\documentclass[12pt]{article}
\usepackage{amsmath,amssymb,amsfonts,graphicx}

\begin{document}
\title{Search for hidden particles in intensity frontier experiment SHiP}

\author{Volodymyr M. Gorkavenko\\
 \it \small ${}^{2}$Department of Physics, Taras Shevchenko National
 University of Kyiv,\\ \it \small 64 Volodymyrs'ka str., Kyiv
 01601, Ukraine}
 \date{}

\maketitle

\setcounter{page}{1}%

\begin{abstract}
Despite the undeniable success of the Standard Model of particle
physics (SM)
 there are some phenomena (neutrino
oscillations, baryon asymmetry of the Universe, dark matter, etc.)
that SM cannot explain. These phenomena indicate that the SM has to
be modified. Most likely there are new particles beyond the SM.
There are many experiments to search for new physics that can be
divided into two types: energy and intensity frontier.
 In experiments of the first type, one tries to directly produce and detect new heavy particles.
 In experiments of the second type, one tries to directly produce and detect new light particles that feebly interact with SM particles. Future intensity  frontier SHiP experiment
  (\textbf{S}earch for \textbf{Hi}dden \textbf{P}articles) at the CERN SPS  is discussed.
  Its advantages and technical characteristics are given.
 \end{abstract}

\section{Introduction}

The Standard Model of particle physics (SM) \cite{SM}  was created
in the mid-1970s. It is one of the greatest successes of physics. It
is experimentally tested with high precision for the processes of
electroweak and strong interactions with the participation of
elementary particles up to energy scale $\sim$100 GeV and for
individual processes up to several TeV. It predicted a number of
particles, last of them (Higgs boson) have been observed in 2012.
However,  the SM cannot explain several phenomena in particle
physics, astrophysics and cosmology. Namely: the SM does not provide
the dark matter candidate; the SM does not explain neutrino
oscillations and the baryon asymmetry of the Universe; the SM cannot
solve the strong CP problem in particle physics, the primordial
perturbations problem and the horizon problem
in cosmology, etc. 

The presence of the problems in the SM indicates the incompleteness
of the Standard Model and the existence of as yet "hidden" sectors
with particles of new physics. Although it may seem surprising, some
of the above-mentioned SM problems really can be solved with help
either heavy or light new particles. Neutrino oscillation and the
smallness of the active neutrino mass can be explained as with help
of new particles with sub-eV mass as well as with help of heavy
particles of the GUT scale, see, e.g., \cite{Strumia}. The same can
be said about the baryon asymmetry of the Universe problem and dark
matter problem: physics on the very different scales can be
responsible for it, see, e.g.,  \cite{kolb}.

\begin{figure}[t]
\includegraphics[width=15cm]{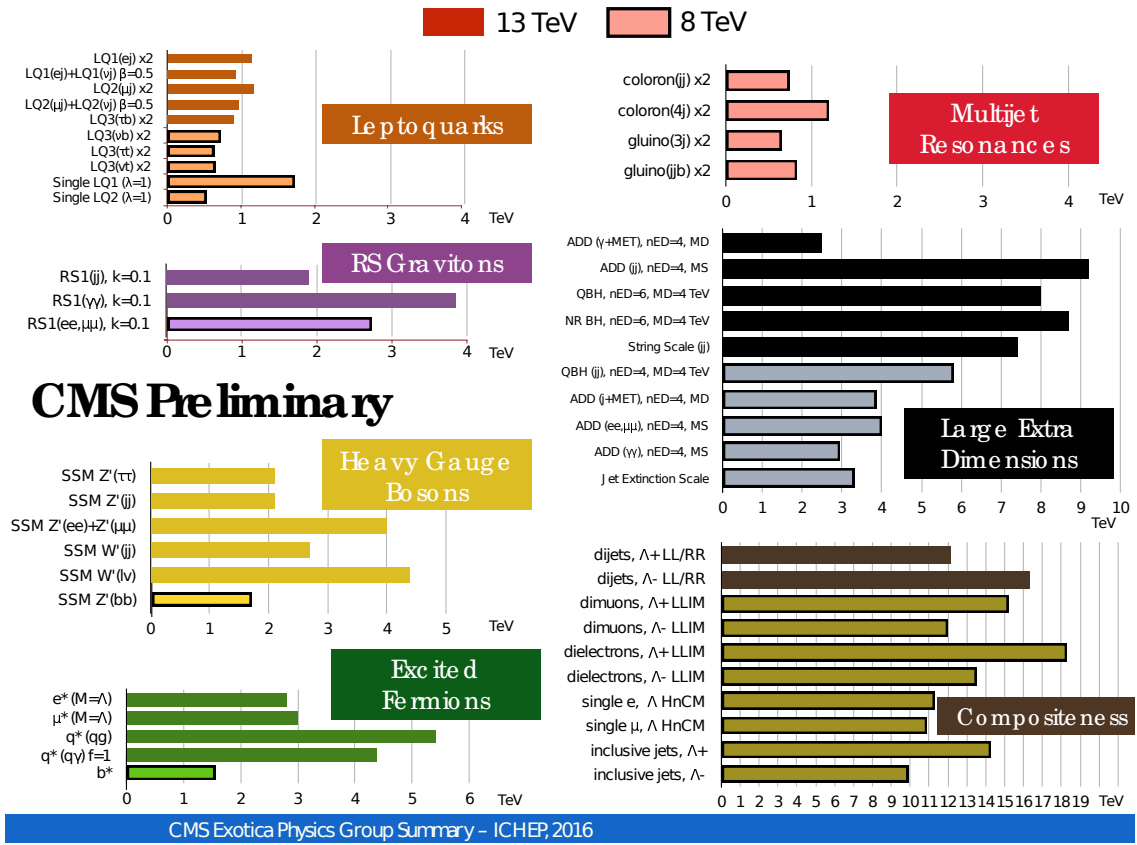}
\caption{Search of new beyond SM particles with mass at TeV region
at CERN CMS. Figure is taken from {\tiny
$https://twiki.cern.ch/twiki/pub/CMSPublic/PhysicsResultsCombined/exo-limits\_ICHEP\_2016.pdf$}}
\end{figure}

Can the new light particles exist in the SM extensions? The answer
is positive. There are many theories beyond SM that have light
particles in the spectrum (e.g., GUT, SUSY, theories with extra
dimensions), see, e.g., \cite{GUT}. Light particles in these
theories can be, e.g., (pseudo)-Goldstone bosons that were produced
as a result of spontaneously symmetry breaking of some not exact
symmetry.  Alternatively, particles can be massless at tree level but it can obtain light mass as a result of loops corrections.

So, it can be two possible answers on the question "why do we not
observe the particles of new physics?"  At first, new particles can
be very heavy (e.g. with mass $M_X\gtrsim 100$ TeV), so it can not
be directly produced at the present-day powerful accelerators like
LHC. On the other hand, new particles can be light (with a mass
bellow or of the order of the electroweak scale) and feebly interact
with particles of the SM (otherwise we would have already seen them
in the experiments). In this case, light new particles can be
produced at many high-energy experiments  but they were not still
observed due to the despite rare events of their production and
complexity of their detection.

Based on the above, there are two types of particle search experiments.

The first of them is energy frontier experiments like LHC or
Fermilab.  In these experiments, one tries to directly produce and
detect new heavy particles assuming that the coupling of new
particles to the SM particles is not very small. New particles with
a mass of several TeV are actively searched in such experiments, see
Fig.~1. Last decades a lot of attention was paid to the energy
frontier experiments.

The second of them is intensity frontier experiments.  In these
experiments, we try to search particles that feebly interact with
the SM particles. So, in intensity frontier experiments we search
for very rare events.  For the successful production of the hidden
particles (to compensate its feeble interaction), this experiment
must have the largest possible luminosity. In this sense beam-dump
experiments are good as intensity frontier experiments for the
search of the GeV-scale hidden particles because their luminosities
are several orders of magnitude larger than at colliders.  Detection
of hidden particles is possible only due to observing their decays
into the SM particles.  So, these experiments must be background
free. Because of feeble interaction with the SM particles, one can
expect its small decay width and long lifetime (here we suppose that
hidden particle does not decay in non-SM channels or corresponding
partial decay width very small). So the detector has to be placed as
far as possible from the point of the hidden particle production.

Intensity frontier experiments have been paid much less attention in
recent years. These experiments include PS 191 (early 1980s), CHARM
(1980s), NuTeV (1990s), DONUT (late 1990s -- early 2000). However,
as it was shown in \cite{Alekhin,fermilab}, the search for new
physics in the region of the mas below the electroweak scale is not
sufficiently investigated.

The difference between energy and intensity frontier experiments for
search hidden particles can be schematically illustrated with the
help of Fig.~2.

In this paper we consider  future intensity  frontier SHiP
(\textbf{S}earch for \textbf{Hi}dden \textbf{P}articles)  beam-dump
experiment in CERN Super Proton Synchrotron (SPS) accelerator. Its
advantages and technical characteristics will be considered. The
class of theories that can be tested on SHiP will be discussed.

\begin{figure}[t]
\includegraphics[width=15cm]{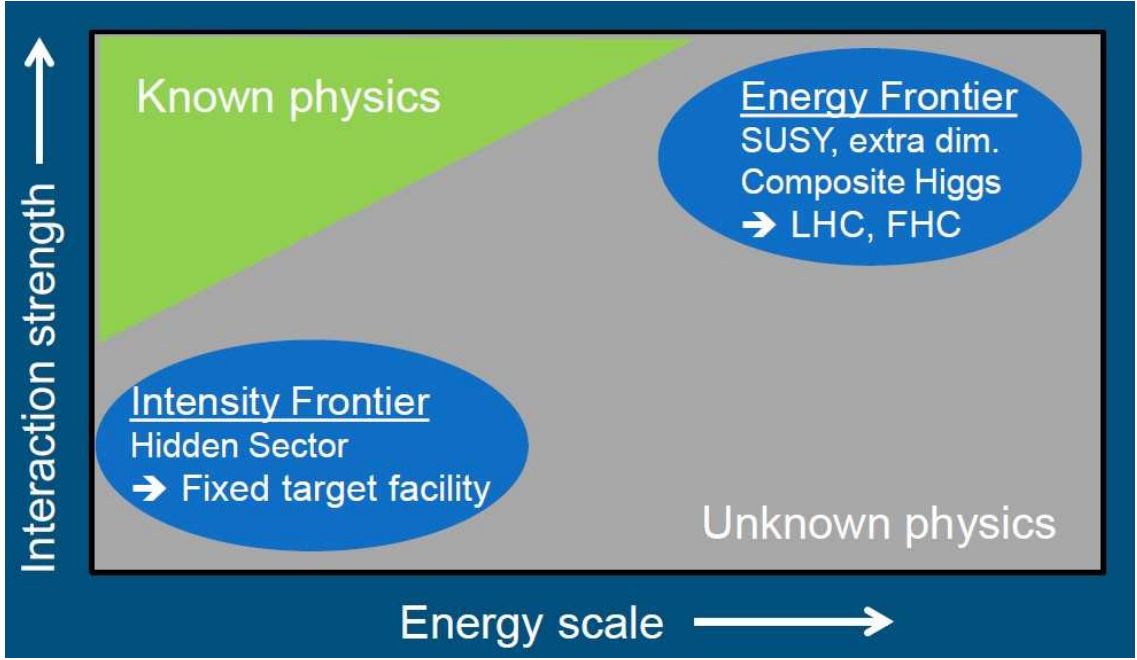}
\vskip-3mm\caption{Different strategy for hidden particles search in
energy and intensity frontier experiments. Figure is taken from
\cite{Alekhin}.}
\end{figure}

\section{Interaction of new particles with the SM particles. Portals}

If we will focus on detecting new light particles, we have
understood that these particles can originate from a big number of
beyond SM theories that predict different parameters for it (mass of
new particles and their coupling to the SM particles). In
particular, such relatively light particles can be mediators due to
interaction as with particles of the SM and very heavy particles of
"hidden sectors". This light particles can be coupled to the
Standard Model sectors either via renormalizable interactions with
small dimensionless couplings ("portals") or by higher-dimensional
operators suppressed by the dimensionful couplings $ \Lambda^{-n} $
corresponding to a new energy scale of the hidden sector
\cite{Alekhin}.

Because of a limited number of  possible types of the particle
(scalar, pseudoscalar, vector, pseudovector, fermion) there is a
limited number of possible effective Lagrangians of interaction such
particles with the SM particles  that satisfy  conditions of Lorentz
and gauge invariance.

Renormalizable portals can be classified into the following 3 types:

\textbf{Vector portal:} new particles are vector  Abelian fields ($A_\mu'$) with the field strength
$F_{\mu\nu}'$, that couple to the hypercharge field $F^{\mu\nu}_Y$ of the SM as
\begin{equation}
  \label{ship_introduction:eq:3}
  \mathcal{L}_\text{Vector portal} = \epsilon F'_{\mu\nu} F^{\mu\nu}_Y,
\end{equation}
where $\epsilon$ is a dimensionless coupling characterizing the
mixing between new vector field with field of the $Z$-boson and the
photon.

\textbf{Scalar portal:} new particles are neutral singlet scalars, $S_i$ that couple to
the  Higgs field
 \begin{equation}\label{darkscalar}
\mathcal{L}_\text{Scalar portal} =  (\lambda_i S_i + g_i S_i^2) (H^{\dagger} H),
\end{equation}
where $\lambda_i$ are dimensionless couplings and  $g_i$ are  couplings with dimension of mass.

\textbf{Neutrino portal:}  new particles are neutral singlet
fermions $N_I$
\begin{equation}\label{sh1}
\mathcal L_\text{Neutrino portal}= F_{\alpha I}\bar L_\alpha \tilde H
N_{I},
\end{equation}
where index  $\alpha=e,\mu,\tau$  corresponds to the lepton
flavors,  $L_{\alpha}$ is for the
lepton doublet, $F_{\alpha I}$ is for the new
matrix of the Yukawa constants,  ${\tilde H}=i\sigma_2H^*$.

Non-renormalizable couplings of new particles to the SM operators
are also possible. For example, pseudo-scalar axion-like particles $A$ couple to  SM as
\begin{equation}
 \mathcal{L}_\text{A}= \sum_{f } \frac{C_{Af}}{2 \, f_a} \bar{f} \gamma^\mu \gamma^5 f \, \partial_\mu A - \frac{\alpha}{8\pi}\, \frac{C_{A\gamma}}{f_a}\, F_{\mu\nu} \tilde{F}^{\mu\nu} \, A - \frac{\alpha_3}{8\pi}\, \frac{C_{A3}}{f_a}\, G_{\mu\nu}^b \tilde{G}^{b\,\mu\nu} \, A,
\end{equation}
where $f =\{\text{quarks,\,leptons,\,neutrinos} \}$,
$F_{\mu\nu}$ is the el\-ec\-tro\-magnetic  field strength tensor, $G_{\mu\nu}^b $ the field strength for the strong force,  the dual field strength tensors are defined as $\tilde{Q}^{\mu\nu} = \frac{1}{2} \epsilon^{\mu\nu\rho\sigma} Q_{\rho\sigma}$.

Another important example is a Chern-Simons like gauge interaction \cite{anomaly} of new pseudo-vector $X_\mu$ particle
\begin{align}
        \mathcal{L}_1&=\frac{C_Y}{\Lambda_Y^2}\cdot X_\mu (\mathfrak D_\nu H)^\dagger H B_{\lambda\rho} \cdot\epsilon^{\mu\nu\lambda\rho}+h.c.\label{gorkavenko:L2b} \\
          \mathcal{L}_2&=\frac{C_{SU(2)}}{\Lambda_{SU(2)}^2}\cdot X_\mu (\mathfrak D_\nu H)^\dagger F_{\lambda\rho} H\cdot\epsilon^{\mu\nu\lambda\rho}+h.c.,\label{gorkavenko:L2a}
\end{align}
where the $\Lambda_Y$, $\Lambda_{SU(2)}$ are new scales of the theory, $C_Y$, $C_{SU(2)}$ are new dimensionless coupling constants,  $B_{\mu\nu}$, $F_{\mu\nu}$ -- field strength tensors of the $U_Y(1)$ and $SU_W(2)$ gauge fields. After spontaneously symmetry breaking of the Higgs field this interaction is effectively reduced to the renormalizable interaction of form
\begin{equation}\label{Lcs}
     \mathcal{L}_\text{CS}=c_z \epsilon^{\mu\nu\lambda\rho} X_\mu Z_\nu \partial_\lambda Z_\rho +c_\gamma \epsilon^{\mu\nu\lambda\rho} X_\mu Z_\nu \partial_\lambda A_\rho+c_w \epsilon^{\mu\nu\lambda\rho} X_\mu W_\nu^- \partial_\lambda W_\rho^+.
\end{equation}

So, from the experimental point of view, one has to test all of the
above mentioned possible new interactions in the wide range of new
particle masses and couplings.

\section{SHiP experiment}

\begin{figure*}[t]
\includegraphics[width=15cm]{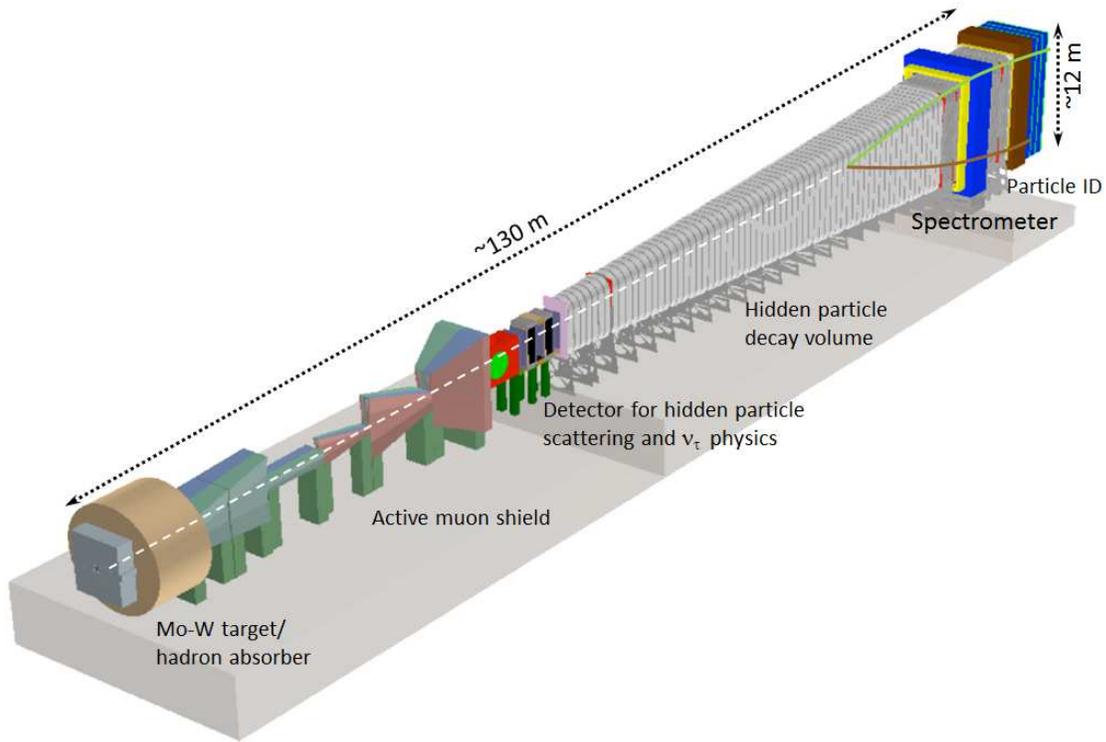}
\caption{General scheme of the SHiP facility. Figure is taken from
\cite{Ahdida1}.}
\end{figure*}

SHiP experiment was first proposed in 2013 \cite{proposed}. The
technical proposal was presented in 2015 \cite{tech}. Theoretical
background, main channels of production and decay of new particles,
preliminary estimations for sensitivity region for different portals
for the SHiP experiment were considered in 2016 \cite{Alekhin}.
Somewhat later, works of a clarifying and complementary nature were
published also \cite{shield,Ahdida,Ahdida1,bond}. Currently, SHiP
collaboration \cite{ship} includes nearly 250 scientist from 53
institutions. The experiment will begin its work allegedly in the
2026 year \cite{Beacham}.

The main goal of the future  SHiP beam-dump experiment in CERN SPS
accelerator is to search new physics in the region of feebly
interacting long-lived light particles including Heavy Neutral
Leptons (HNL), vector, scalar, axion portals to the Hidden Sector,
and light supersymmetric particles. The experiment provides great
opportunities for the study of neutrino physics as well.

Now, we describe the work of the SHiP experiment, see Fig.~3. A
beamline from the CERN SPS accelerator will transmit 400 GeV protons
at the SHiP. The proton beam will strike in a Molybdenum and
Tungsten fixed target at a center-of-mass energy $E_{CM} \approx 27$
GeV. Approximately $2\cdot10^{20}$ proton-target collisions are
expected  in 5 years of SHiP operation. The great number of light
the SM particles and hadrons will be produced under such collisions.
Hidden  particles are expected to be predominantly produced in the
decays of the hadrons.

The main concept of SHiP functioning is the following. Almost all
the produced SM particle should be either absorbed or deflected in
the magnetic field (muons). Remaining events with SM particles can
be rejected using specially developed cuts.
 If hidden particles will decay into the SM particles inside the decay volume, the last will be detected. It will mean the existence of the hidden particles.

So, the target will be followed by a 5m long iron hadron absorber.
It will absorb the hadrons and the electromagnetic radiation from
the target, but decays of the mesons result in a large flux of muons
and neutrinos. After the hadron stopper is located the system of
shielding magnets (which extends over a length of  $\sim 40$m) to
deflect muons away from the fiducial decay volume \cite{shield}.

Despite the aim to search long-living particles, the sensitive
volume should be situated as close as possible to the proton target
due to the relatively large transverse momentum of the hidden
particles with respect to the beam axis. The minimum distance is
determined by necessity for the system to absorb the electromagnetic
radiation and hadrons producing from the proton-target collisions
and to reduce the beam-induced muon flux.

The system of detectors of the SHiP consists of the two parts. Just
after the hadron absorber and muon shield is located detector system
for recoil signatures of hidden sector particle scattering and for
neutrino physics. Neutrino detector has a mass of nearly 10 tons.
Investigation of the neutrino physics is based on a hybrid detector
similar to the detector of the OPERA Collaboration \cite{opera}.
Besides the study of the neutrino physics this system allows us to
detect and veto charged particles produced outside the main decay
volume.

The second detector system consists of the fiducial decay volume
that is contained in nearly  50 m long rectangular vacuum tank. In
order to suppress the background from neutrinos interacting in the
fiducial volume, it is maintained at a pressure of $O(10^{-3})$ bar.
The decay volume is surrounded by background taggers to tag neutrino
and muon inelastic scattering in the surrounding structures, which
may produce long-lived neutral Standard Model particles whose decay
can mimic signal events. The vacuum tank followed a large
spectrometer with a rectangular acceptance of 5m in width and 10m in
height and calorimeter. The system is constructed in such a way to
detect as many final states as possible in order to be sensitive to
a very wide range of models that can be tested. With the help of
Tabl.~1 one can see what modification of the SM is tested depending
on final states of the hidden particles decay.

\begin{table}[th] \noindent\caption{Modification of the SM that
can be tested on SHiP depending on final states hidden particles
decay.}\vskip3mm\tabcolsep4.5pt

\begin{center}
\noindent{\footnotesize\begin{tabular}{|l|l|l|} \hline
\multicolumn{1}{|c}{\rule{0pt}{5mm}Decay modes} &
\multicolumn{1}{|c}{\rule{0pt}{5mm}Final states} &
\multicolumn{1}{|c|}{\rule{0pt}{5mm}Models tested} \\[2mm]%
\hline%
{\rule{0pt}{5mm}meson and lepton} &  $\pi l$, $K l$, $l$ & $\nu$ portal, HNL,\\%
 & $(l=e,\mu,\tau)$ & SUSY neutralino\\
two leptons & $e^+\,e^-$, $\mu^+ \mu^-$ & V, S and A portals,\\%
 & & SUSY  s-goldstino\\
two mesons & $\pi^+\,\pi^-$, $K^+\,K^-$  & V, S and A portals,\\%
 & & SUSY  s-goldstino\\
3 body & $l^+\,l^-\,\nu$ & HNL, neutralino \\[2mm]%
\hline
\end{tabular}}
\end{center}

\end{table}

It should be noted that SHiP experiment gives great opportunities for study of neutrino physics.
In result of nearly $2\cdot 10^{20}$ proton-target collisions it is expected production of
$N_{\nu_\tau}=5.7\cdot10^{15}$ $\nu_{\tau}$ and $\nu_{\bar\tau}$ neutrino, $N_{\nu_e} = 5.7\cdot10^{18}$ electron neutrino,
 and $N_{\nu_{\mu}} = 3.7\cdot10^{17}$  muon neutrino.  It is expected to detect nearly $10^4$ $\tau$-neutrino and at first
 detect anti $\tau$-neutrino. It is very important  because of until now only 14 $\tau$-neutrino candidates by
 experiment DONUT in Fermilab and 10  $\tau$-neutrino
candidates by experiment OPERA in CERN. No event with anti
$\tau$-neutrino was still observed.

\section{Conclusions}

There are some indisputable phenomena that point to the fact that SM
has to be modified and complemented by a new particle (particles).
We are sure that there is new physics, but we do not know where to
search for it. There are many theoretical possibilities to modify
the SM by a scalar, pseudoscalar, vector, pseudovector or fermion
particles of new physics. These particles may be sufficiently heavy
the electroweak scale and scale of the energy of the present
colliders. But these particles may be light (with mass less then
electroweak scale) and feebly interact with the SM particles. The
main task now is to experimentally observe particles of new physics.

Since the possibilities for increasing energy of the present
colliders are limited by high costs and heavy new particles are
difficult to produce, it seems reasonable to check another variant
and try to find light particles of new physics in intensity frontier
experiments.

The goal of the SHiP experiment is to search new physics in the
region of feebly interacting long-lived light particles including
HNL, vector, scalar, axion particles with mass $\le 10$ GeV. There
are theoretical predictions for the sensitivity region of the SHiP
experiment for each type of new physics particles (in the mass new
particles and coupling constant coordinates). The experiment will
provide great opportunities for the study of neutrino physics also.

Since the idea of searching new light feebly interacting particles
is very tempting and promising there are other projects such as
REDTOP at the PS beamlines, NA64++,
 NA62++, LDMX, AWAKE,  KLEVER at the  SPS beamlines,  FASER, MATHUSLA,
CODEX-b at the LHC. All these experiments are compared and
summarized in \cite{Beacham}. It is possible that great discoveries
in particle physics are right ahead.

\vskip3mm \textit{The work was presented at the conference "New
trends in high energy physics", May 12-18, Odessa, Ukraine. I also
thank Kyrylo Bondarenko for  useful discussion and helpful
comments.}


\newpage




\begin{thebibliography}{20}                                                                                                %
\bibitem{SM}  S.L. Glashow,  Nucl. Phys. \textbf{22}, 579 (1961);
S. Weinberg,  Phys. Rev. Lett. \textbf{19}, 1264 (1967); A. Salam,
in Proc. of 8th Nobel Symposium (Ed. N Svartholm) (Stockholm:
Almquist and Wiksells, 1968) p. 367.

\bibitem{Strumia} A. Strumia, F. Vissani,  arXiv:hep-ph/0606054 (2010).

\bibitem{kolb} D.S. Gorbunov and V.A. Rubakov, \textit{Introduction to the Theory of the Early Universe:
Hot Big Bang Theory} (World Scientific, 2011).

\bibitem{GUT} W. de Boer,  Prog. Part. Nucl. Phys. \textbf{33}, 201 (1994).

\bibitem{Alekhin} S. Alekhin {\it et al.}, Rept. Prog. Phys. \textbf{79},  124201 (2016).

\bibitem{anomaly} I. Antoniadis, A. Boyarsky, S. Espahbodi,
O. Ruchayskiy, J.D. Wells,  Nucl. Phys. B\textbf{824}, 296 (2010).

\bibitem{fermilab} J. Alexander {\it et al.},  arXiv:1608.08632, FERMILAB-CONF-16-421 (2016).

\bibitem{proposed} W. Bonivento {\it et al.},  arXiv:1310.1762  (2013).

\bibitem{tech} M. Anelli {\it et al.},   arXiv:1504.04956, CERN-SPSC-2015-016, SPSC-P-350  (2015).

\bibitem{shield} A. Akmete {\it et al.},  JINST \textbf{12}, P05011 (2017).

\bibitem{Ahdida} C. Ahdida {\it et al.},  J. High Energ. Phys. 2019: 77  (2019).

\bibitem{Ahdida1} C. Ahdida {\it et al.},  JINST \textbf{14},  P03025 (2019).

\bibitem{bond} I. Boiarska, K. Bondarenko, A. Boyarsky,
V. Gorkavenko, M. Ovchynnikov, A. Sokolenko,  arXiv:1904.10447v2
(2019).

\bibitem{ship} https://ship.web.cern.ch/ship

\bibitem{Beacham} J. Beacham, et al.,  arXiv:1901.09966, CERN-PBC-REPORT-2018-007 (2019).

\bibitem{opera} OPERA collaboration,
JINST \textbf{4}, P04018 (2009).


\end{thebibliography}
\end{document}